\documentclass[11pt]{article}
\usepackage{amsmath}
\usepackage{amssymb}
\usepackage{bbm}
\usepackage{epsfig}
\usepackage{feynmf}
\allowdisplaybreaks[4]
\begin{document}
\begin{fmffile}{fmftempl}

\title{Flavour changing strong interaction effects on top quark physics at the
LHC}
\author{P.M. Ferreira$^{1,}$~\footnote{ferreira@cii.fc.ul.pt},
O. Oliveira$^{2,}$~\footnote{orlando@teor.fis.uc.pt} and 
R. Santos$^{1,}$~\footnote{rsantos@cii.fc.ul.pt}\\ 
$^1$ Centro de F\'{\i}sica Te\'orica e Computacional, Faculdade de Ci\^encias,\\
Universidade de Lisboa, Av. Prof. Gama Pinto, 2, 1649-003 Lisboa, Portugal \\
$^2$ Centro de F\'{\i}sica Computacional, Universidade de Coimbra, Portugal}
\date{September, 2005} 
\maketitle
\noindent
{\bf Abstract.} We perform a model independent analysis of the flavour changing 
strong interaction vertices relevant to the LHC. In particular, the 
contribution of dimension six operators to single top production in various 
production processes is discussed, together with possible hints for identifying 
signals and setting bounds on physics beyond the standard model.
\vspace{1cm}

\section{Introduction}

The top quark is a fascinating particle. It is the heaviest particle discovered 
so far, and one of the least known~\cite{rev}. These two facts make it an ideal 
laboratory for searches of physics beyond the Standard Model (SM). The 
theoretical implications of new physics may be studied in two ways: by 
proposing a new theory and computing its effects on observable quantities; or by
parameterizing all possible effects of new physics using effective operators of
dimensions higher than four, and analysing their impact on measurable variables.
The full set of dimension five and six operators was obtained by 
Buchm\"uller and Wyler~\cite{buch}. There has been much work on the effect of 
these operators on the properties of the top quark. For instance, the authors
of the series of papers listed in reference~\cite{whis1, whis2, whis3} studied 
how several types of dimension five and six operators affected the top quark, 
both at the Tevatron and at the LHC, as well as other types of physics (such as 
$b\bar{b}$ production). The authors of~\cite{saav} undertook a detailed study of
the $W t b$ vertex at LHC conditions. New physics contributions for flavour 
changing neutral currents were studied in~\cite{fcnc}. The contributions 
from four-fermion operators to $t\bar{t}$ production were considered in 
ref.~\cite{4f}.  

In this work we will focus on the case of dimension six operators with flavour
changing interactions involving the top quark, and the processes for which
one has, at the LHC, a single quark $t$ being produced. The set of operators we
chose includes, after symmetry breaking, the dimension five ones mentioned 
above, but also new contributions not considered simultaneously before. Our
methodology will also be slightly different from that of previous works in this 
area: whenever possible, we will present full analytical expressions for our
results, so that our experimental colleagues at the Tevatron or LHC may use them
in their Monte Carlo generators, to study the sensibility of the experiments to
this new physics. The outline of this paper is as follows: in 
section~\ref{sec:op} we will describe the operator set we have chosen and the 
reasons that motivated that choice. We
will discuss some subtleties related to the number of independent operators and 
the inclusion of four-fermion terms, as well. In section~\ref{sec:eff} we will 
compute the contributions arising from these operators to the top quark width 
and also to a number of physical processes of single top production at the LHC.
Analysis of these results and conclusions will be presented in 
section~\ref{sec:conc}. 

\section{Dimension six operators}
\label{sec:op}

Following the philosophy and notation of ref.~\cite{buch}, we consider the 
lagrangean
\begin{equation}
{\cal L} \;\;=\;\; {\cal L}^{SM} \;+\; \frac{1}{\Lambda}\,{\cal L}^{(5)} \;+\;
\frac{1}{\Lambda^2}\,{\cal L}^{(6)} \;+\; O\,\left(\frac{1}{\Lambda^3}\right)
\;\;\; ,
\label{eq:l}
\end{equation}
where ${\cal L}^{SM}$ is the SM lagrangean, $\Lambda$ is the energy scale at 
which new physics makes itself manifest and ${\cal L}^{(5)}$ and 
${\cal L}^{(6)}$ are operators of dimension 5 and 6 respectively which, like 
${\cal L}^{SM}$, are invariant under the SM gauge group and built with its 
fields. Imposing baryon and lepton number conservation, the term 
${\cal L}^{(5)}$ is eliminated. However, after spontaneous symmetry breaking, 
dimension five operators will appear in the lagrangean, arising from 
${\cal L}^{(6)}$. The total number of dimension six operators is quite large 
(see~\cite{buch} for the full list). It is clearly not practical to consider 
them all when studying the effects of new physics, and so some selection 
criteria are needed. We assumed from the start that, whatever new physics 
contributions appear at the lagrangean~\eqref{eq:l} as dimension six operators, 
low-energy phenomenology will not be substantially altered by them. By 
``low-energy" we mean any process occurring below the LEP energies. Another 
selection criterion was that we are only interested in operators involving a 
single top quark - we have in mind the study of single top quark production, so 
operators with more than a top quark are of no interest to us. Finally, we 
restricted ourselves to a particular type of new physics: flavour changing 
vertices involving the top quark and the strong interactions, namely, vertices 
of the form $g\,t\,\bar{c}$ or $g\,t\,\bar{u}$. It is a choice like any other, 
but motivated by two arguments: first, the LHC environment is ideally suited to 
study the strong sector of the SM (much like LEP was designed to study its 
electroweak sector) so, if a deviation from normal QCD vertices exists, it will 
likely be possible to discover it at the LHC; second, some of the operators we 
will be considering - the so-called chromomagnetic operators - are generated in
the context of the SM at high orders, but their effects are expected to be too 
small to be measured. However, they appear in more general theories, such as 
supersymmetric models or theories with multiple Higgs doublets (see, for 
example,~\cite{chro}).

These criteria reduce the number of operators involving gluons to just two, 
namely
\begin{align}
{\cal O}_{uG} &= \;\;i\,\frac{\alpha_{ij}}{\Lambda^2}\,\left(\bar{u}^i_R\,
\lambda^a\, \gamma^\mu\,D^\nu\,u^j_R\right)\,G^a_{\mu\nu} \nonumber 
\vspace{0.2cm} \\
{\cal O}_{uG\phi} &= \;\;\frac{\beta_{ij}}{\Lambda^2}\,\left(\bar{q}^i_L\,
\lambda^a\, \sigma^{\mu\nu}\,u^j_R\right)\,\tilde{\phi}\,G^a_{\mu\nu} \;\;\; ,
\label{eq:op}
\end{align} 
where $u^i_R$ is an up quark right spinor (an SU(2) singlet, therefore), $q^i_L$
a left quark doublet spinor, $\tilde{\phi}$ the charge conjugate of the Higgs 
doublet, $G^a_{\mu\nu}$ the gluon tensor. $\alpha_{ij}$ and $\beta_{ij}$ are 
complex couplings and the remaining quantities are well-known. The indices 
$(i\,,\,j)$ on the spinors are flavour indices, indicating to which generation 
each one belongs. As an example of the application of our criteria, consider 
another {\em a priori} possible operator from~\cite{buch}, 
\begin{equation}
{\cal O}_{qG} \;\;=\;\; i\,\left(\bar{q}^i_L\,\lambda^a\,\gamma^\mu\,D^\nu\,
q^j_L \right)\,G^a_{\mu\nu} \;\;\; .
\end{equation}
In terms of up and down quarks this operator becomes
\begin{equation}
{\cal O}_{qG}\;\;=\;\; i\,\left(\bar{u}^i_L\,\lambda^a\,\gamma^\mu\,D^\nu\,u^j_L
\;+\; \bar{d}^i_L\,\lambda^a\,\gamma^\mu\,D^\nu\,d^j_L\right)\,G^a_{\mu\nu} 
\;\;\; .
\end{equation}
The second term of this operator gives an interaction involving only down type
quarks. It would therefore have direct implications on low-energy physics (such
as bottom physics, since one of the doublets, per one of our criteria, must 
belong to the third generation). As such, its size is already immensely 
constrained by the wealth of existing data. Hence, the relevance of this 
operator to new phenomena will be very limited and we do not consider it. It
might also seem bizarre that we are considering the second of the operators of
equation~\eqref{eq:op}, as it has a Higgs field in it, and our criteria did not 
include phenomenology of scalar fields. However, after spontaneous symmetry
breaking, the neutral component of $\phi$ acquires a non-zero vev ($\phi_0\,
\rightarrow\,\phi_0\,+\,v$, with $v\,=\,246/\sqrt{2}$ GeV) and the operator 
decomposes into several pieces containing Higgs scalars and one without them, 
namely
\begin{equation}
{\cal O}_{uG\phi} \;\;\rightarrow\;\;  \beta_{ij}\,\frac{v}{\Lambda^2}\,\left(
\bar{u}^i_L\,\lambda^a\,\sigma^{\mu\nu}\,u^j_R\right)\,G^a_{\mu\nu}
\;\;\; .
\end{equation}
The dimension 6 operator thus becomes identical to a dimension 5 chromomagnetic
one.

One of the generation indices on the operators~\eqref{eq:op} must correspond to
the third family, and so our lagrangean for new physics becomes
\begin{align}
{\cal L}\;\; =&\;\;\; \alpha_{tu}\,{\cal O}_{tu}\;+\; \alpha_{ut}\,{\cal O}_{ut}
\;+\; \beta_{tu}\,{\cal O}_{tu\phi}\;+\;\beta_{ut}\,{\cal O}_{ut\phi}\;+\;
\mbox{h.c.} \nonumber \vspace{0.2cm} \\
 =& \;\;\;\frac{i}{\Lambda^2}\,\left[\alpha_{tu}\,\left(\bar{t}_R\,\lambda^a\,
\gamma^\mu \,D^\nu\,u_R\right)\;+\;  \alpha_{ut}\,\left(\bar{u}_R\,\lambda^a\, 
\gamma^\mu\, D^\nu\,t_R\right)\right]\,G^a_{\mu\nu} \;\;\;+ \nonumber 
\vspace{0.2cm} \\
 & \;\;\;\frac{v}{\Lambda^2}\,\left[\beta_{tu}\,\left(\bar{t}_L\,\lambda^a\,
\sigma^{\mu\nu}\,u_R\right)\;+\; \beta_{ut}\,\left(\bar{u}_L\,\lambda^a\,
\sigma^{\mu\nu}\,t_R\right)\right]\,G^a_{\mu\nu} \;\;+\;\; \mbox{h.c.}
\;\;\;, 
\end{align}
for the vertices $g\,\bar{t}\,u$, $g\,\bar{u}\,t$, and an analogous
lagrangean (with different coupling constants, clearly) for $g\,\bar{t}\,c$ and
$g\,t\,\bar{c}$.  Then, the Feynman rules for the anomalous vertices $g\,\bar{t}
\,u$ and $g\,t\,\bar{u}$ are~\footnote{Gluon momenta are defined as incoming,
quark momenta follow the sense of the arrows in the figure.} 
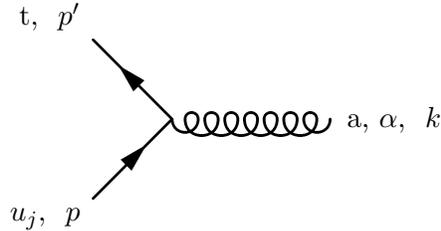
\vspace{0.5cm}
\begin{eqnarray}
 & &
\parbox{50mm}{\begin{fmfchar*}(100,60)
  \fmfleft{qp,qa}
  \fmf{fermion}{qp,gv,qa}
  \fmfright{g}
  \fmf{gluon}{gv,g}
  \fmflabel{$u_j$, $~p$}{qp}
  \fmflabel{t, $~p'$}{qa}
  \fmflabel{a, $\alpha$, $ ~ k$}{g}
\end{fmfchar*} \vspace{0.8cm}}
   \nonumber \\ & &
   \frac{\lambda^a}{\Lambda^2}
   \Bigg\{ \gamma_\mu \gamma_R \left( \alpha_{tj} p_\nu ~ + ~
                                      \alpha^*_{jt} p'_\nu
                                              \right) ~ + ~
           v \, \sigma_{\mu\nu} \left( \beta_{tj} \gamma_R ~ + ~
                                       \beta^*_{jt} \gamma_L
                                              \right)
             \Bigg\} \times \nonumber \\
  & & \hspace{7cm}
      \Big( k^\mu g^{\nu \alpha} ~ - ~ k^\nu g^{\mu \alpha} \Big)
\nonumber
\end{eqnarray}
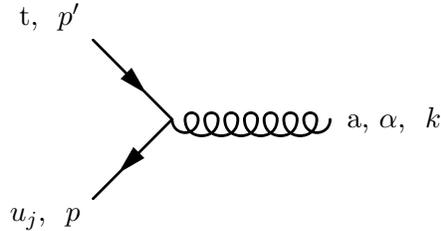
\vspace{0.5cm}
\begin{eqnarray}
 & &
\parbox{50mm}{\begin{fmfchar*}(100,60)
  \fmfleft{qp,qa}
  \fmf{fermion}{qa,gv,qp}
  \fmfright{g}
  \fmf{gluon}{gv,g}
  \fmflabel{$u_j$, $~p$}{qp}
  \fmflabel{t, $~p'$}{qa}
  \fmflabel{a, $\alpha$, $ ~ k$}{g}
\end{fmfchar*} \vspace{0.8cm}}
   \nonumber \\ & &
   \frac{\lambda^a}{\Lambda^2}
   \Bigg\{ \gamma_\mu \gamma_R \left( \alpha_{jt} p_\nu ~ + ~
                                      \alpha^*_{tj} p'_\nu
                                              \right) ~ + ~
           v \, \sigma_{\mu\nu} \left( \beta_{jt} \gamma_R ~ + ~
                                       \beta^*_{tj} \gamma_L
                                              \right)
             \Bigg\} \times \nonumber \\
  & & \hspace{7cm}
      \Big( k^\mu g^{\nu \alpha} ~ - ~ k^\nu g^{\mu \alpha} \Big) 
\end{eqnarray}
\vspace{0.5cm}

If one uses the fermionic equations of motion and integrates by parts, one 
obtains the following relations between operators:
\begin{align}
{\cal O}^{\dagger}_{ut} &= {\cal O}_{tu}\;-\;\frac{i}{2} (\Gamma^{\dagger}_u\,
{\cal O}^{\dagger}_{u t \phi} \,+\, \Gamma_u \,{\cal O}_{t u \phi}) \nonumber \\
{\cal O}^{\dagger}_{ut} &= {\cal O}_{tu}\;-\;i\, g_s\, \bar{t}\, \gamma_{\mu}\,
\gamma_R\, \lambda^a\,u\, \sum_i  (\bar{u}^i\, \gamma^{\mu}\, \gamma_R\, 
\lambda_a u^i\,+\, \bar{d}^i\, \gamma^{\mu}\, \gamma_R\, \lambda_a\, d^i)
\;\;\; ,
\label{eq:rel}
\end{align}
where $\Gamma_u$ are the Yukawa coupling matrices associated with the $u$ quarks
and $g_s$ is the strong gauge coupling. In the second of these equations we see 
the appearance of four-fermion operators. Then, as long as one considers these
extra contributions, we obtain two relations between ${\cal O}_{tu}$, 
${\cal O}_{ut}$, ${\cal O}_{tu\phi}$ and ${\cal O}_{ut\phi}$ and so we can set
two of the $\{\alpha\,,\,\beta\}$ constants to zero. If, however, we 
leave out the four fermion contributions, only the first of eqs.~\eqref{eq:rel} 
is valid and therefore only one of $\{\alpha_{tu}\,,\,\alpha_{ut}\,,\,\beta_{tu}
\,,\,\beta_{ut}\}$ may be set to zero. These considerations are important for 
more elaborate calculations than those that will be undertaken in this paper -
for instance, the process $g\,g\,\rightarrow\,\bar{t}\,c$. In the present work,
for reasons to be understood later, the inclusion of the four-fermion operators
shown in eq.~\eqref{eq:rel} has no effect on the results and thus we do not 
consider them. Therefore, we use only one of the equations~\eqref{eq:rel} and 
reduce the dependence of our expressions to three complex parameters per 
generation, setting $\beta_{tu}$ and $\beta_{tc}$ to zero. Nevertheless, we will
present our full results, for completeness. Finally, equation~\eqref{eq:rel} 
shows the advantage of having chosen this particular operator set. With it, we 
may take advantage of the gauge invariance of the theory, which is gives us 
relations between the several operators, allowing us to reduce the number of 
independent parameters.  

\section{Effects of the new physics on top quark observables}
\label{sec:eff}

The vertices $g\,\bar{t}\,c$ and $g\,\bar{t}\,u$ constitute new possible decay 
channels for the top quark. Hence they will contribute to the top's width, and
a straightforward calculation shows that
\begin{align}
\Gamma (t \rightarrow u g) &=\;  \frac{m^3_t}{12 \pi\Lambda^4}\,\Bigg\{ m^2_t 
\,\left|\alpha_{ut}  +  \alpha^*_{tu}\right|^2 \,+\, 16 \,v^2\, \left(\left| 
\beta_{tu} \right|^2 + \left| \beta_{ut} \right|^2 \right) \;\;\; +
\vspace{0.3cm} \nonumber \\
 & \hspace{2.2cm}\, 8\, v\, m_t\,\mbox{Im}\left[ (\alpha_{ut}  + \alpha^*_{tu}) 
\, \beta_{tu} \right] \Bigg\}
\label{eq:wid}
\end{align}
where we assumed that all of the quark masses, except the top's, are zero. 
Having performed the full calculation, we verified that this is an excellent 
approximation. This will allow us to impose limits on the values of the new 
couplings. The main decay channel for the top quark is $t\,\rightarrow\,b\,W$,
the tree level decay width being given, in the SM, by
\begin{equation}
 \Gamma (t \rightarrow b W ) \; =\; \frac{g^2}{64\,\pi}\,\left|V_{tb}\right|^2\,
\frac{m^3_t}{M^2_W}\,\left(1\,+\,\frac{M^2_W}{m^2_t}\,-\,2\frac{M^4_W}{m^4_t} 
\right)\, \left( 1\, -\, \frac{M^2_W}{m^2_t} \right) \;\;\; ,
\end{equation}
assuming $m_b\,=\,0$. After QCD corrections and using $m_t\,=\,175$ GeV, this 
becomes~\cite{rev,qcdc}
\begin{equation}
 \Gamma (t \rightarrow b W ) \; =\; 1.42\,\left|V_{tb}\right|^2\;\;\; 
\mbox{GeV}\;\;\; .
\end{equation}
Taking $V_{tb}\,\sim\,1$ this will be, to very good approximation, the total 
width of the top quark. Then,
\begin{align}
\frac{\Gamma (t \rightarrow u g)}{\Gamma_t} &=\;\;
\frac{10^{-3}}{\Lambda^4}\,\Bigg\{ 3.07\,\left|\alpha_{ut}+
\alpha^*_{tu}\right|^2 \,+\, 48.53\, \left(\left|\beta_{tu} \right|^2 + \left|
\beta_{ut} \right|^2 \right) \;\;\;+ \vspace{0.3cm} \nonumber \\
 & \hspace{2cm} 24.41 \mbox{Im} \left[ (\alpha_{ut}  +  \alpha^*_{tu}) \, 
\beta_{tu} \right] \Bigg\}\;\;\; ,
\end{align}
with $\Lambda$ expressed in TeV. Let us now assume that these flavour changing 
decays are not observed at the LHC. If $L$ is the degree of precision with which
the top quark width will be known at the LHC then we will have $\Gamma (t 
\rightarrow q g)/\Gamma_t\;\leq\;L$, where $q\,=\,u\,,\,c$. Since
$\beta_{tu}$ and $\beta_{tc}$ were set to zero, upper bounds on the $\{\alpha\,,
\,\beta\}$ couplings are obtained as $\left(|\alpha_{qt}\,+\,\alpha_{tq}^*|
\right)/\Lambda^2\,\leq\,18.05\,\sqrt{L}$ $\mbox{TeV}^{-2}$ and $|\beta_{qt}|/
\Lambda^2\,\leq\,4.51\,\sqrt{L}$ $\mbox{TeV}^{-2}$. Assuming that it will be 
possible to measure $\Gamma_t$ with 10\% precision at the LHC (which seems 
reasonable, considering preliminary studies already done~\cite{topw}), these 
upper limits become, respectively, 5.71 and 1.44 $\mbox{TeV}^{-2}$. The authors
of ref.~\cite{whis2} used CDF data~\cite{cdf} to impose the limit 
$BR(t\,\rightarrow\,c\,g)\,<\,0.45$ and obtaining from this $|\beta_{ct}|/
\Lambda^2\,\leq\,3.1$ $\mbox{TeV}^{-2}$, which is in agreement with the result
we obtained above.

\begin{figure}[htb]
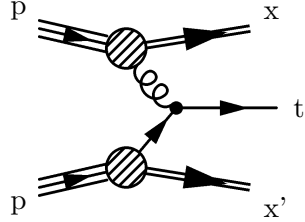

\begin{center}
\parbox{50mm}{\begin{fmfchar*}(100,60)
  \fmfleft{p2,p1}
  \fmfright{x2,t,x1}
  \fmf{fermion,tension=2}{p1,v1}
  \fmf{fermion,tension=2}{p2,v2}
  \fmf{heavy}{v1,x1}
  \fmf{heavy}{v2,x2}
  \fmfblob{.15w}{v1}
  \fmfblob{.15w}{v2}
  \fmf{gluon}{v1,va}
  \fmf{fermion}{v2,va,t}
  \fmfdot{va}
  \fmflabel{p}{p1}
  \fmflabel{p}{p2}
  \fmflabel{x}{x1}
  \fmflabel{x'}{x2}
  \fmflabel{t}{t}
  \fmffreeze
  \renewcommand{\P}[3]{\fmfi{plain}{vpath(__#1,__#2) shifted (thick*3(#3))}}
  \P{p1}{v1}{0,0.5}
  \P{p1}{v1}{0,-0.5}
  \P{p2}{v2}{0,0.5}
  \P{p2}{v2}{0,-0.5}
\end{fmfchar*}}
\vspace{0.5cm}
\caption{Direct top production at the LHC via flavour changing gluon vertices}
\label{fig:dir}
\end{center}
\end{figure}

The flavour changing vertices $g\,\bar{t}\,c$, $g\,\bar{t}\,u$, $\ldots$ induce 
several new channels for single top production. The first, and 
as we will see, the most important, is direct top production, in which a gluon 
and a quark u/c are extracted from the colliding protons, as is illustrated in
fig.~\ref{fig:dir}. With a single particle in the final state and the colliding 
protons having large parton density functions (pdf's), this channel is 
favoured over all others. Due to the one-particle final state, the cross section
calculation is very simple, and we obtain 
\begin{align}
\sigma(p\,p\, \rightarrow\, t)   &=  \sum_{q\,=\,u,c}\,
\frac{\pi \, m_t}{12\,\Lambda^4}\,\left\{\,m^2_t\,|\alpha_{qt}+ \alpha^*_{tq}|^2
\,+\,16\,v^2\,\left(|\beta_{qt}|^2 + |\beta_{tq}|^2 \right)\;\;\;+\right.
\vspace{0.3cm}\nonumber \\
 &\left. \;\;\;8\,v\,m_t\,\mbox{Im}\,\left[(\alpha_{qt}+\alpha^*_{tq})\,
\beta_{tq} \right]\right\} \;\int^1_{\frac{m^2_t}{E_{CM}^2}}\frac{2\, m_t}{
E_{CM}^2 \, x_1} f_g (x_1)\, f_q (m^2_t/(E_{CM}^2\, x_1)) \, dx_1 
\end{align}
where $E_{CM}$ is the proton-proton center of mass energy (14 TeV for the LHC), 
$f_g$ and $f_q$ are the parton density functions for the gluon and the quark $q$
(an up or charm quark). We see that the partonic cross section factorizes from
the pdf integration and, using the results from eq.~\eqref{eq:wid}, we obtain
\begin{equation}
\sigma(p\,p\,\rightarrow\,t)\;\;=\;\;\sum_{q\,=\,u,c}\,\Gamma (t\,\rightarrow\,q
\,g)\,
\frac{\pi^2}{m_t^2}\;\int^1_{m^2_t/E_{CM}^2}\frac{2\, m_t}{E_{CM}^2\, x_1} f_g 
(x_1)\, f_q (m^2_t/(E_{CM}^2\, x_1)) \, dx_1 \;\;\; .
\label{eq:ppt}
\end{equation}
If we assume that the branching ratio $BR(t\,\rightarrow\,b\,W)$ is 
approximately 100\% then, using the above results for the top width at the SM, 
we may express $\Gamma (t\,\rightarrow\,q\,g)$ as $1.42\,\left|V_{tb}\right|^2\,
BR(t\,\rightarrow\,q\,g)$. We numerically performed the integration in 
eq.~\eqref{eq:ppt} using the CTEQ6M parton density functions~\cite{cteq6} 
setting the factorization scale equal to $m_t$~\footnote{This choice of 
factorization scale produces smaller values for the cross section than setting
$\mu_F$ equal to the partonic center-of-mass energy (see, for instance, 
reference~\cite{singt}). For comparison between SM and anomalous cross sections,
though, this has no bearing on our conclusions.}. The total cross section is
\begin{equation}
\sigma(p\,p\,\rightarrow\,t)\;\;=\;\;\left[10.5\,BR(t\,\rightarrow\,u\,g)\;+\;
1.7\,BR(t\,\rightarrow\,c\,g)\right]\,\left|V_{tb}\right|^2\,10^4\;\;\;\mbox{pb}
\;\;\;.
\label{eq:bru}
\end{equation}
The bigger contribution from the $u$ quark channel stems from the fact the pdf
for that quark is larger than the $c$ quark's. For anti-top production, we 
obtain
\begin{equation}
\sigma(p\,p\,\rightarrow\,\bar{t})\;\;=\;\;\left[2.8\,BR(\bar{t}\,\rightarrow\,
\bar{u}\,g)\;+\;1.7\,BR(\bar{t}\,\rightarrow\,\bar{c}\,g)\right]\,\left|V_{tb}
\right|^2\,10^4\;\;\;\mbox{pb} \;\;\;.
\label{eq:brc}
\end{equation}
In terms of the couplings $\alpha$ and $\beta$, the cross sections become
\begin{align}
\sigma^{(u)}(p\,p\,\rightarrow\,t) &=\;\;\frac{1}{\Lambda^4}\,\Bigg\{321\,
\left|\alpha_{ut} + \alpha^*_{tu}\right|^2\,+\,5080\,\left(\left|\beta_{tu}
\right|^2 + \left| \beta_{ut} \right|^2\right)\;\;\;+\vspace{0.3cm}\nonumber \\
 & \hspace{1.6cm}2556\,\mbox{Im}\left[( \alpha_{ut} + \alpha^*_{tu})\, 
\beta_{tu} \right] \Bigg\}\;\;\mbox{pb}
\label{eq:sigu}
\end{align}
for top production via an $u$ quark, and, for the $c$ quark contribution,
\begin{align}
\sigma^{(c)}(p\,p\,\rightarrow\,t) &=\;\;\frac{1}{\Lambda^4}\,\Bigg\{50\,
\left|\alpha_{ct} + \alpha^*_{tc}\right|^2\,+\,796\,\left(\left|\beta_{tc}
\right|^2 + \left| \beta_{ct} \right|^2\right)\;\;\;+\vspace{0.3cm}\nonumber \\
 & \hspace{1.6cm} 400\,\mbox{Im}\left[( \alpha_{ct} + \alpha^*_{tc}) \, 
\beta_{tc} \right] \Bigg\}\;\;\mbox{pb} \;\;\; .
\label{eq:sigc}
\end{align}
Using the expressions~\eqref{eq:bru} and~\eqref{eq:brc} one arrives, in a 
straightforward manner, to the expressions for anti-top production cross 
sections. A crude upper bound on the single top cross sections may be obtained 
if we consider the most unfavourable case in which the decays $t\,\rightarrow\,u
\,g$, etc, have no visible effect on the top width. Using the estimates of the 
previous paragraphs and considering both branching ratios identical, the cross 
section is bounded as $[\sigma(p\,p\,\rightarrow\,t)\,+\,\sigma(p\,p\,
\rightarrow\,\bar{t})]\,\leq\, 1.7\times 10^5\,L$ pb. With 5\% precision on the 
measurement of the top's width at the LHC, this would give an upper bound of 
8350 pb for the cross section of this process. Single top quark production has 
been extensively studied~\cite{topcr}. The value we obtained above is quite 
large, if compared with the expected value for single top production at the LHC 
at NLO within the SM framework, roughly $319.7\,\pm\,19.3$ pb~\cite{singt}. 
Alternatively, we can ask ourselves what the values of the couplings $\{\alpha\,
 ,\,\beta\}$ should be so that direct top production via flavour changing gluon 
interactions is not observed. This means that the anomalous contribution is such
that $\sigma_{AN} \leq \mbox{Max}\{\Delta_T\,,\, \Delta_E\}$, where $\Delta_T$ 
and $\Delta_E$ are the theoretical and experimental uncertainties on the cross 
section. Bounds on the couplings are then obtained considering each one 
separately and, using the above mentioned value $\Delta_T\,=\,19.3$ pb, we get 
(in units of TeV$^{-2}$)
\begin{align}
\frac{\left|\alpha_{ut} + \alpha^*_{tu}\right|}{\Lambda^2}\,\leq\,\mbox{Max}\,
\left(0.22\,,\,0.049\,\sqrt{\Delta_E}\right)\;\;\;, & 
\;\;\; \frac{\left|\beta_{ut}\right|}{\Lambda^2}\;\leq\;\mbox{Max}\,\left(0.055
\,,\,0.013\, \sqrt{\Delta_E}\right) \vspace{0.3cm} \nonumber \\
\frac{\left|\alpha_{ct} + \alpha^*_{tc}\right|}{\Lambda^2}\,\leq\,\mbox{Max}\,
\left(0.44\,, \,0.10\,\sqrt{\Delta_E}\right)   \;\;\;, &
\;\;\; \frac{\left|\beta_{ct}\right|}{\Lambda^2}\;\leq\;\mbox{Max}\,\left(0.11
\,,\,0.025\, \sqrt{\Delta_E}\right)\;\;\;.
\end{align}
The expected experimental error is $\Delta_E\,=\,16$ pb (5\% of the total cross
section~\cite{topw}), therefore the current theoretical error dominates. These 
results are to be compared with the limits predicted in ref.~\cite{whis3} for 
the $\beta$ couplings: from a study involving the simulation of the possible LHC
backgrounds and for 10 $fb^{-1}$ of data, they obtained, when converted into our
notation, $|\beta_{ut}|/\Lambda^2\,\leq\,0.021$ $\mbox{TeV}^{-2}$ and 
$|\beta_{ct}|/\Lambda^2\,\leq\,0.046$ $\mbox{TeV}^{-2}$.

\begin{figure}[htb]
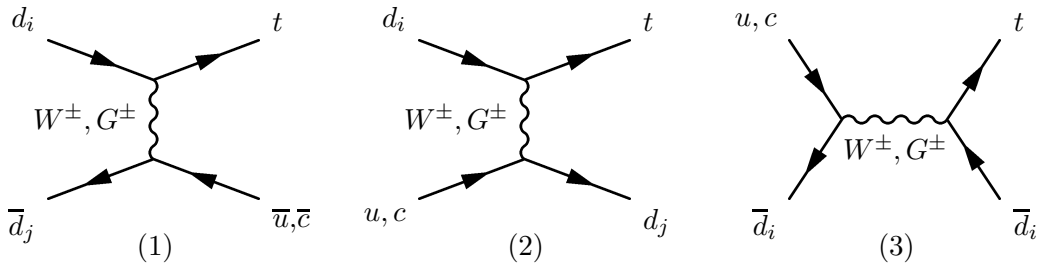

\begin{center}
\begin{minipage}[t]{.3\textwidth}
\parbox{50mm}{
\begin{fmfchar*}(100,60)
  \fmfleft{d1,d2}
  \fmfright{u,t}
  \fmf{fermion}{d2,v1,t}
  \fmf{fermion}{u,v2,d1}
  \fmf{photon,label=$W^{\pm},,G^{\pm}$}{v1,v2}
  \fmflabel{${\overline d}_j$}{d1}
  \fmflabel{${\overline u}$,${\overline c}$}{u}
  \fmflabel{$t$}{t}
  \fmflabel{$d_i$}{d2}
\end{fmfchar*}
}

\begin{center}
\hspace{-1.25cm}(1)
\end{center}
\end{minipage}
\begin{minipage}[t]{.3\textwidth}
\parbox{50mm}{\begin{fmfchar*}(100,60)
  \fmfleft{d1,d2}
  \fmfright{u,t}
  \fmf{fermion}{d2,v1,t}
  \fmf{fermion}{d1,v2,u}
  \fmf{photon,label=$W^{\pm},,G^{\pm}$}{v1,v2}
  \fmflabel{$u,c$}{d1}
  \fmflabel{$d_j$}{u}
  \fmflabel{$t$}{t}
  \fmflabel{$d_i$}{d2}
\end{fmfchar*}}

\begin{center}
\hspace{-1.25cm}(2)
\end{center}
\end{minipage}
\begin{minipage}[t]{.3\textwidth}
\parbox{50mm}{\begin{fmfchar*}(100,60)
  \fmfleft{d1,d2}
  \fmfright{u,t}
  \fmf{fermion}{d2,v1,d1}
  \fmf{fermion}{u,v2,t}
  \fmf{photon,label=$W^{\pm},,G^{\pm}$}{v1,v2}
  \fmflabel{${\overline d}_i$}{d1}
  \fmflabel{${\overline d}_i$}{u}
  \fmflabel{$t$}{t}
  \fmflabel{$u,c$}{d2}
\end{fmfchar*}}

\begin{center}
\hspace{-1.25cm}(3)
\end{center}
\end{minipage}
\vspace{0.5cm}
\caption{$t$ and $s$ channels for single top production according to the
Standard Model, at tree-level.}
\label{fig:smst}
\end{center}
\end{figure}

\begin{figure}[htb]
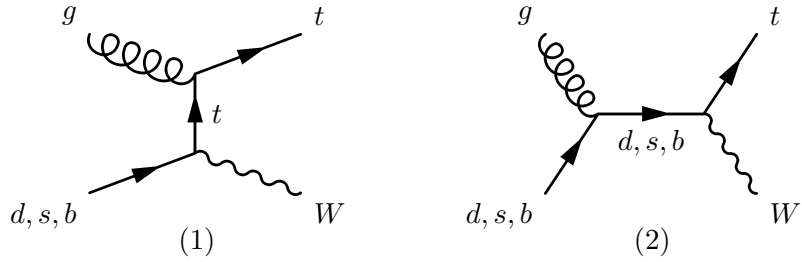

\begin{center}
\begin{minipage}[t]{.3\textwidth}
\parbox{50mm}{\begin{fmfchar*}(100,60)
  \fmfleft{d1,d2}
  \fmfright{u,t}
  \fmf{gluon}{d2,v1}
  \fmf{fermion}{d1,v2}
  \fmf{fermion,label=$t$}{v2,v1}
  \fmf{fermion}{v1,t}
  \fmf{photon}{v2,u}
  \fmflabel{$d,s,b$}{d1}
  \fmflabel{$W$}{u}
  \fmflabel{$t$}{t}
  \fmflabel{$g$}{d2}
\end{fmfchar*}}

\begin{center}
\hspace{-1.25cm}(1)
\end{center}
\end{minipage}
\hspace{1cm}
\begin{minipage}[t]{.3\textwidth}
\parbox{50mm}{\begin{fmfchar*}(100,60)
  \fmfleft{d1,d2}
  \fmfright{u,t}
  \fmf{gluon}{d2,v1}
  \fmf{fermion}{d1,v1}
  \fmf{fermion,label=$d,,s,,b$}{v1,v2}
  \fmf{fermion}{v2,t}
  \fmf{photon}{v2,u}
  \fmflabel{$d,s,b$}{d1}
  \fmflabel{$W$}{u}
  \fmflabel{$t$}{t}
  \fmflabel{$g$}{d2}
\end{fmfchar*}}

\begin{center}
\hspace{-1.25cm}(2)
\end{center}
\end{minipage}
\caption{Associated single top production according to the Standard Model, at
tree-level.}
\label{fig:smtw}
\end{center}
\end{figure}

All observables we have considered thus far depend on $\Lambda^{-4}$. For a 
large energy scale, then, the effects of new physics will be very small. There
are however processes for which the scale dependence is $\Lambda^{-2}$, namely
the interference terms between the operators~\eqref{eq:op} and the SM process
for which a single top quark is produced. In the SM at tree-level, the Feynman
diagrams for single top production are shown in figures~\ref{fig:smst} 
and~\ref{fig:smtw}, for the reactions $p\,p\,\rightarrow\,t\,j$ and $p\,p\,
\rightarrow\,t\,W$. Most of these diagrams will be suppressed by elements of the
CKM matrix, and also by small pdf contributions from the incoming quarks. Using 
the CTEQ6M structure functions and considering all possible combinations of 
incoming partons we obtain, for the tree-level cross sections, the values listed
in table 1. The SM results are different from those quoted in 
references~\cite{singt}, though not by much. This is due to the size of the QCD 
corrections and also to the relatively low factorization scale we chose. It is 
easy, with our anomalous vertices, to build diagrams that 
will interfere with the SM processes; namely, those of figures~\ref{fig:int1} 
and~\ref{fig:int2}. Notice how, due to the standard gluon vertex, the down-type 
quarks always belong to the same generation. The interference with the SM 
diagrams of fig.~\ref{fig:smst} will thus be restricted to that case. The 
calculation of the interference terms gives, for the differential partonic cross
sections, ($q\,=\,u\,,\,c$)
\begin{align}
\frac{d\,\sigma^{tj}_1}{dt}\;&= \; K\,\frac{s\,+\,t\,-\,m_t^2}{s^2\,(t\,-\,
m_W^2)}\,\,\frac{\mbox{Re}(\beta{qt})}{\Lambda^2} \nonumber 
\vspace{0.3cm} \\
\frac{d\,\sigma^{tj}_2}{dt}\;&= \; \frac{K}{s\,(s\,+\,t\,+\,m_W^2\,-\,
m_t^2)}\,\,\frac{\mbox{Re}(\beta{qt})}{\Lambda^2} \nonumber 
\vspace{0.3cm} \\ 
\frac{d\,\sigma^{tj}_3}{dt}\;&= \; K\,\frac{s\,+\,t\,-\,m_t^2}{s^2\,(s\,-\,
m_W^2)}\,\, \frac{\mbox{Re}(\beta{qt})}{\Lambda^2}\nonumber 
\vspace{0.3cm} \\
\frac{d\,\sigma^{tW}_1}{dt}\;&= \; \frac{3\,K}{32\,m_W^2}\,
\frac{(t\,-\,2 m_W^2)\,\left[(t\,-\,m_W^2)\,m_t^2\,+\,(m_W^2\,-\,2 s\,-\,t)\,t
\right]}{t\,(t\,-\,m_t^2)\,s^2}\,\,\frac{\mbox{Re}(\beta{qt})}{
\Lambda^2} \nonumber \vspace{0.3cm} \\
\frac{d\,\sigma^{tW}_2}{dt}\;&= \; \frac{3\,K}{32\,m_W^2}\,\frac{2\,m_W^4\,
-\,t\,(m_W^2\,-\,2 s\,-\,t)\,+\,2\,m_t^2\,(t\,-\,2 m_W^2)}{t\,s^2}\,\,
\frac{\mbox{Re}(\beta{qt})}{\Lambda^2} \;\;\; ,
\label{eq:int}
\end{align}
where $K$ is a factor including all couplings, spin and colour factors and other
constants, given by~\footnote{We left out the conversion factor $0.389\times
10^9$ pb TeV$^2$}
\begin{equation}
K\;=\; \frac{8}{9}\,\sqrt{\frac{2\,\alpha_S}{\pi}}\,G_F\,m_t\,m_W^2\,v\,|V_{qi}
\,V_{ti}| \;\;\;,
\end{equation}
and we have used, for the CKM matrix elements $V_{ij}$, the central values of
the PDG data~\cite{pdg}. Notice how these results depend solely on one of the 
couplings - this is due to the particular chiral structure of the anomalous 
operators~\eqref{eq:op}, but stems also from the fact that we are considering 
all quark masses to be zero, except the top's. If we relax this approximation 
(which means, in effect, taking into account the diagrams in fig.~\ref{fig:smst}
with Goldstone bosons $G^\pm$), terms in $\alpha_{ut}$, $\alpha_{tu}$ and 
$\beta_{tu}$ will appear in~\eqref{eq:int}. However, direct calculation of those
terms shows that they are much smaller than those shown in~\eqref{eq:int} (at 
least 50 times smaller, but typically much inferior to that). It is interesting 
to observe that these particular processes probe one single anomalous coupling, 
thus allowing us, in theory, to determine it independently of the others. 
Summing all possible contributions from the incoming partons and performing the 
integrations on both $t$ and the pdf's, we obtain the results shown in 
table 1. As we can see, the anomalous cross sections are typically 
much smaller than the expected SM values. The largest deviation from the SM 
predictions occurs for the production of a top and bottom quarks, for which we 
have
\begin{equation}  
\frac{\sigma^{AN}(p\,p\,\rightarrow\,t\,b)\,+\,\sigma^{AN}(p\,p\,\rightarrow\,
\bar{t}\,b)}{\sigma^{SM}(p\,p\,\rightarrow\,t\,b)\,+\,\sigma^{SM}(p\,p\,
\rightarrow\,\bar{t}\,b)}\;\;=\;\; \left[ 1.0\,\frac{\mbox{Re}(\beta_{ut})}{
\Lambda^2} \,+\, 2.3\,\frac{\mbox{Re}(\beta_{ct})}{\Lambda^2}\right]\;\% 
\;\;\; .
\end{equation}
Even in this more favourable case, we see that the deviation from the SM is at
the most a few percent of the values of the couplings. We would expect that QCD
corrections to these calculations would not affect significantly this ratio, as 
both the SM and anomalous cross sections should have similar contributions from 
higher order corrections. The reason for the smallness of these numbers is easy 
to understand: the diagrams for which one has interference are suppressed by 
off-diagonal CKM matrix elements. There are two ways for this suppression to 
occur: for instance, the process $d\,\bar{d}\,\rightarrow\,t\, \bar{u}$ is 
suppressed quite heavily by the CKM product $V_{td}\,V_{ud}$ - even though 
$V_{ud}$ is a diagonal CKM element, $V_{td}$ represents a two family ``jump" and
as such is miniscule. 
The second possibility is, for example, the process $u\,s\,\rightarrow\,t\,s$, 
which is suppressed by $V_{ts}\,V_{us}$ - the product of these two CKM elements,
each of them representing a one family ``jump", is also extremely small. 
\begin{table}[t]
\begin{center}
\begin{tabular}{ccc}  \hline\hline \\
 Process & SM cross section (pb) & Anomalous cross section (pb)  \\ & & \\ 
\hline \\ $p\,p\,\rightarrow\,t\,j$ & 151.0 & 
$\displaystyle{0.80\,\frac{\mbox{Re}(\beta_{ut})}{\Lambda^2}\,+\, 0.10\,
\frac{\mbox{Re}(\beta_{ct})}{\Lambda^2}}$ \\ 
 & & \\ 
 $p\,p\,\rightarrow\,t\,b$ & 4.8 & 
$\displaystyle{0.07\,\frac{\mbox{Re}(\beta_{ut})}{\Lambda^2}\,+\,0.09\,
\frac{\mbox{Re}(\beta_{ct})}{\Lambda^2}}$ \\ 
 & & \\ 
 $p\,p\,\rightarrow\,\bar{t}\,j$ & 89.7 &
$\displaystyle{0.08\,\frac{\mbox{Re}(\beta_{ut})}{\Lambda^2}\,+\,0.08\,
\frac{\mbox{Re}(\beta_{ct})}{\Lambda^2}}$ \\
 & & \\ 
 $p\,p\,\rightarrow\,\bar{t}\,b$ & 3.0 &
$\displaystyle{0.01\,\frac{\mbox{Re}(\beta_{ut})}{\Lambda^2}\,+\,0.09\,
\frac{\mbox{Re}(\beta_{ct})}{\Lambda^2}}$ \\
 & & \\ 
 $p\,p\,\rightarrow\,t\,W$ & 31.1 &
$\displaystyle{0.04\,\frac{\mbox{Re}(\beta_{ut})}{\Lambda^2}\,+\,0.08\,
\frac{\mbox{Re}(\beta_{ct})}{\Lambda^2}}$ \\
 & & \\ 
 $p\,p\,\rightarrow\,\bar{t}\,W$ & 31.1 & 
$\displaystyle{0.03\,\frac{\mbox{Re}(\beta_{ut})}{\Lambda^2}\,+\,0.08\,
\frac{\mbox{Re}(\beta_{ct})}{\Lambda^2}}$\\
 & & \\\hline\hline\hline 
\label{tab:res}
\end{tabular}
\caption{Results for the interference cross sections. The processes $p\,p\,
\rightarrow\,t\,j$, $p\,p\,\rightarrow\,\bar{t}\,j$ refer to production of 
non-bottom quarks alongside the top.}
\end{center}
\end{table}

As for the four-fermion operators, besides those shown in 
equation~\eqref{eq:rel}, there are {\em a priori} two other types that we could
consider in this calculation~\cite{buch}. Using the same criteria applied to the
gluonic operators, we can classify them as:
\begin{itemize}
\item{Type 1,
\begin{equation}
{\cal O}_{u_1}\;\;=\;\; \frac{g_s\,\gamma_{u_1}}{\Lambda^2}
\left(\bar{t}\, \lambda^a\,\gamma^{\mu}\, \gamma_R\, u\right)\,\left(\bar{q}
\, \lambda^a\,\gamma_{\mu}\, \gamma_R\, q\right)\;+\;\mbox{h.c.} \;\;\; ,
\end{equation}
where $q$ is any given quark, other than the top; these are the operators that
appear in equation~\eqref{eq:rel}.}
\item{Type 2,
\begin{equation}
{\cal O}_{u_2}\;\;=\;\; \frac{g_s\,\gamma_{u_2}}{\Lambda^2}
\left[\left(\bar{t}\, \lambda^a\, \gamma_L\, u^\prime\right)\,\left(
\bar{u}^{\prime\prime}\,\lambda^a\, \gamma_R\, u\right) \; + \; \left(\bar{t}\,
\lambda^a\, \gamma_L\, d^\prime\right)\,\left(\bar{d}^{\prime\prime}\,\lambda^a
\,\gamma_R\, u\right) \right] \;+\;\mbox{h.c.} \;\;\; ,
\end{equation}
with down and up quarks from several possible generations, excluding the top
once more;}
\item{Type 3,
\begin{equation}
{\cal O}_{u_3}\;\;=\;\; \frac{g_s\,\gamma_{u_3}}{\Lambda^2}
\left[\left(\bar{t}\, \lambda^a\, \gamma_R\, u\right)\,\left(
\bar{b}\,\lambda^a\, \gamma_R\, d^\prime\right) \; - \; \left(\bar{t}\,
\lambda^a\, \gamma_R\, d^\prime\right)\,\left(\bar{b}\,\lambda^a\,\gamma_R\,u
\right) \right] \;+\;\mbox{h.c.} \;\;\; ,
\label{eq:ga31}
\end{equation}
and also,
\begin{equation}
\frac{g_s\,\gamma_{u_3}^*}{\Lambda^2}
\left[\left(\bar{t}\, \lambda^a\, \gamma_L\, u\right)\,\left(
\bar{d}^\prime\,\lambda^a\, \gamma_L\, d^{\prime\prime}\right) \; - \; \left(
\bar{t}\, \lambda^a\, \gamma_L\, d\right)\,\left(\bar{d}^\prime\,\lambda^a\,
\gamma_L\,u^{\prime\prime}\right) \right] \;+\;\mbox{h.c.} \;\;\; .
\label{eq:ga32}
\end{equation}
}
\end{itemize}
All of these operators could, in principle, interfere with the SM processes of 
single top production shown in fig.~\ref{fig:smst}. However, due to their chiral
structure and the fact that we considered all of the quarks other than the top
to have zero mass, a straightforward calculation shows that their interference
terms with the SM diagrams are zero.

\section{Conclusions}
\label{sec:conc}

In this paper we analysed the simpler processes affected by our choice of 
dimension six operators, which modeled, in an effective way, the possibility
of strong interactions provoking flavour violations involving the top quark.
These induce new decay possibilities for the top quark, namely $t\,\rightarrow
\,c\,g$ or $t\,\rightarrow\,u\,g$. Precision measurements of the top quark width
at the LHC will be able to set stringent bounds on the new anomalous couplings. 
The existence of these new vertices gives rise to a new process for production
of a single top quark - direct top production - a process for which the 
calculated cross section is, {\em a priori}, quite large. At this point, 
detailed simulations of this process at the LHC are necessary to clarify whether
it is possible to distinguish it from other processes of single top production. 
We anticipate a possible complication, the fact that the top quark is produced
with virtually no transverse momentum and a great longitudinal momentum 
component. The results of its decay, which will then interact with the 
detectors, might therefore have also a strong $p_L$ component and concentrate on
the small angle region, making their observation more difficult. 

There are interference terms between the Feynman diagrams that describe single
top production in the SM with two quarks in the initial state and those with a
single anomalous gluon coupling. The resulting cross sections depend on 
$\Lambda^{-2}$, instead of $\Lambda^{-4}$ as in the previous examples. These 
should therefore be a privileged source of bounds on the new operators, 
specially given that these quantities depend only on one of the new couplings. 
However, the values obtained for the interference cross sections are very small
compared with the expected SM results, or indeed with the other anomalous cross
sections obtained, which depend on $\Lambda^{-4}$. The reason for this, we 
found, was a strong CKM suppression. Again, a full simulation of this process 
at the LHC is needed, to investigate whether some kinematical cuts might be 
capable of extricating the interference terms from the remaining cross section,
but that seems unlikely at this stage. The four-fermion operators did not
have any importance to the results presented here. They have no contribution to
the top's decay into a light quark and a gluon, nor to the process of direct top
production. Further, their interference with the SM processes of single top 
production is zero. 

Finally, the interest of the operator set we considered in this work is not
exhausted on the simpler processes of single top production we studied here. 
They have a strong impact on processes such as $g\,g\,\rightarrow\,t\,\bar{u}$,
$g\,u\,\rightarrow\,t\,g$ which, due to the large gluon pdf, one can expect to 
be very significant at the LHC. In ref.~\cite{pap2} we will study these 
processes and there, the four-fermion operators will have an important role to 
play as well.

\vspace{0.25cm}
{\bf Acknowledgments:} Our thanks to Augusto Barroso, Lu\'{\i}s Bento and our
colleagues from LIP for valuable discussions. Our further thanks to A.B. and 
Ant\'onio Onofre for a careful reading of the manuscript. This work is 
supported by Funda\c{c}\~ao para a Ci\^encia e Tecnologia under contract 
PDCT/FP/FNU/50155/2003 and POCI/FIS/59741/2004. P.M.F. is supported by FCT 
under contract SFRH/BPD/5575/2001.

\begin{figure}[htb]
\begin{center}
\begin{minipage}[t]{.3\textwidth}
\parbox{50mm}{\begin{fmfchar*}(100,60)
 \fmfleft{d1,d2}
  \fmfright{u,t}
  \fmf{fermion}{d1,v1,d2}
  \fmf{fermion}{u,v2,t}
  \fmf{gluon,label=$g$,lab.dist=.1w}{v1,v2}
  \fmflabel{${\overline d}_i$}{d1}
  \fmflabel{${\overline u},{\overline c}$}{u}
  \fmflabel{$t$}{t}
  \fmflabel{$d_i$}{d2}
  \fmfdot{v2}
\end{fmfchar*}}

\begin{center}
\hspace{-1.25cm}(1)
\end{center}
\end{minipage}
\begin{minipage}[t]{.3\textwidth}
\parbox{50mm}{\begin{fmfchar*}(100,60)
  \fmfleft{d1,d2}
  \fmfright{t,u}
  \fmf{fermion}{d1,v1,t}
  \fmf{fermion}{d2,v2,u}
  \fmf{gluon,label=$g$}{v1,v2}
  \fmflabel{$d_i$}{d1}
  \fmflabel{$t$}{u}
  \fmflabel{$d_i$}{t}
  \fmflabel{$u,c$}{d2}
  \fmfdot{v2}
\end{fmfchar*}}

\begin{center}
\hspace{-1.25cm}(2)
\end{center}
\end{minipage}
\begin{minipage}[t]{.3\textwidth}
\parbox{50mm}{\begin{fmfchar*}(100,60)
  \fmfleft{d1,d2}
  \fmfright{t,u}
  \fmf{fermion}{t,v1,d1}
  \fmf{fermion}{d2,v2,u}
  \fmf{gluon,label=$g$}{v1,v2}
  \fmflabel{${\overline d}_i$}{d1}
  \fmflabel{$t$}{u}
  \fmflabel{${\overline d}_i$}{t}
  \fmflabel{$u,c$}{d2}
  \fmfdot{v2}
\end{fmfchar*}}

\begin{center}
\hspace{-1.25cm}(3)
\end{center}
\end{minipage}
\vspace{0.5cm}
\caption{Diagrams with an anomalous gluon vertex that interfere with those of
fig.~\ref{fig:smst}.}
\label{fig:int1}
\end{center}
\end{figure}
\begin{figure}[htb]
\begin{center}
\parbox{50mm}{\begin{fmfchar*}(100,60)
  \fmfleft{d1,d2}
  \fmfright{u,t}
  \fmf{gluon}{d2,v1}
  \fmf{fermion}{d1,v2}
  \fmf{fermion,label=$u,,c$}{v2,v1}
  \fmf{fermion}{v1,t}
  \fmf{photon}{v2,u}
  \fmflabel{$d,s,b$}{d1}
  \fmflabel{$W$}{u}
  \fmflabel{$t$}{t}
  \fmflabel{$g$}{d2}
  \fmfdot{v1}
\end{fmfchar*}}
\vspace{0.5cm}
\caption{Diagram with an anomalous gluon vertex that interfere with those of
fig.~\ref{fig:smtw}.}
\label{fig:int2}
\end{center}
\end{figure}

\end{fmffile}
\end{document}